\pdfoutput=1

\documentclass[12pt,a4paper]{article}

\usepackage{ifthen} 
\newboolean{pdflatex}
\setboolean{pdflatex}{true} 

\newboolean{articletitles}
\setboolean{articletitles}{true} 

\newboolean{uprightparticles}
\setboolean{uprightparticles}{false} 

\newboolean{inbibliography}
\setboolean{inbibliography}{false} 

\usepackage{microtype}
\usepackage{lineno}  
\usepackage{xspace} 
\usepackage{caption} 

\usepackage{graphicx}  
\usepackage{color}
\usepackage{colortbl}
\graphicspath{{./figs/}} 

\usepackage{amsmath} 
\usepackage{amssymb}
\usepackage{amsfonts}
\usepackage{upgreek} 

\newcommand*\patchAmsMathEnvironmentForLineno[1]{%
\expandafter\let\csname old#1\expandafter\endcsname\csname #1\endcsname
\expandafter\let\csname oldend#1\expandafter\endcsname\csname
end#1\endcsname
 \renewenvironment{#1}%
   {\linenomath\csname old#1\endcsname}%
   {\csname oldend#1\endcsname\endlinenomath}%
}
\newcommand*\patchBothAmsMathEnvironmentsForLineno[1]{%
  \patchAmsMathEnvironmentForLineno{#1}%
  \patchAmsMathEnvironmentForLineno{#1*}%
}
\AtBeginDocument{%
\patchBothAmsMathEnvironmentsForLineno{equation}%
\patchBothAmsMathEnvironmentsForLineno{align}%
\patchBothAmsMathEnvironmentsForLineno{flalign}%
\patchBothAmsMathEnvironmentsForLineno{alignat}%
\patchBothAmsMathEnvironmentsForLineno{gather}%
\patchBothAmsMathEnvironmentsForLineno{multline}%
\patchBothAmsMathEnvironmentsForLineno{eqnarray}%
}

\usepackage{hyperref}    
\usepackage[all]{hypcap} 

\usepackage{xspace} 
\usepackage{upgreek}

\def\lhcb {\mbox{LHCb}\xspace}

\def\babar  {\mbox{BaBar}\xspace}
\def\belle  {\mbox{Belle}\xspace}

\def\dzero  {\mbox{D0}\xspace}

\def\MagUp {\mbox{\em Mag\kern -0.05em Up}\xspace}

\ifthenelse{\boolean{uprightparticles}}%
{

 \def\Ppi         {\ensuremath{\uppi}\xspace}

 \def\Ppsi        {\ensuremath{\uppsi}\xspace}

 \def\PDelta      {\ensuremath{\Delta}\xspace}                 
 \def\PXi      {\ensuremath{\Xi}\xspace}                 
 \def\PLambda      {\ensuremath{\Lambda}\xspace}                 
 \def\PSigma      {\ensuremath{\Sigma}\xspace}                 
 \def\POmega      {\ensuremath{\Omega}\xspace}                 
 \def\PUpsilon      {\ensuremath{\Upsilon}\xspace}                 
 
 \def\PB      {\ensuremath{\mathrm{B}}\xspace}                 
                  
 \def\PD      {\ensuremath{\mathrm{D}}\xspace}

 \def\PJ      {\ensuremath{\mathrm{J}}\xspace}                 
 \def\PK      {\ensuremath{\mathrm{K}}\xspace}

 \def\Pb      {\ensuremath{\mathrm{b}}\xspace}                 
 \def\Pc      {\ensuremath{\mathrm{c}}\xspace}                 
 \def\Pd      {\ensuremath{\mathrm{d}}\xspace}

 \def\Pi      {\ensuremath{\mathrm{i}}\xspace}

 \def\Pp      {\ensuremath{\mathrm{p}}\xspace}

 \def\Ps      {\ensuremath{\mathrm{s}}\xspace}                 
 \def\Pt      {\ensuremath{\mathrm{t}}\xspace}

}
{

 \def\Ppi         {\ensuremath{\pi}\xspace}

 \def\Ppsi        {\ensuremath{\psi}\xspace}                 
                  
 \mathchardef\PDelta="7101
 \mathchardef\PXi="7104
 \mathchardef\PLambda="7103
 \mathchardef\PSigma="7106
 \mathchardef\POmega="710A
 \mathchardef\PUpsilon="7107
                  
 \def\PB      {\ensuremath{B}\xspace}                 
                  
 \def\PD      {\ensuremath{D}\xspace}

 \def\PJ      {\ensuremath{J}\xspace}                 
 \def\PK      {\ensuremath{K}\xspace}

 \def\Pb      {\ensuremath{b}\xspace}                 
 \def\Pc      {\ensuremath{c}\xspace}                 
 \def\Pd      {\ensuremath{d}\xspace}

 \def\Pi      {\ensuremath{i}\xspace}

 \def\Pp      {\ensuremath{p}\xspace}

 \def\Ps      {\ensuremath{s}\xspace}                 
 \def\Pt      {\ensuremath{t}\xspace}

}

\makeatletter
\ifcase \@ptsize \relax
  \newcommand{\miniscule}{\@setfontsize\miniscule{4}{5}}
\or
  \newcommand{\miniscule}{\@setfontsize\miniscule{5}{6}}
\or
  \newcommand{\miniscule}{\@setfontsize\miniscule{5}{6}}
\fi
\makeatother

\DeclareRobustCommand{\optbar}[1]{\shortstack{{\miniscule (\rule[.5ex]{1.25em}{.18mm})}
  \\ [-.7ex] $#1$}}

\def\dquark    {{\ensuremath{\Pd}}\xspace}

\def\squark    {{\ensuremath{\Ps}}\xspace}

\def\cquark    {{\ensuremath{\Pc}}\xspace}

\def\bquark    {{\ensuremath{\Pb}}\xspace}

\def\tquark    {{\ensuremath{\Pt}}\xspace}

\def\pion   {{\ensuremath{\Ppi}}\xspace}

\def\pip    {{\ensuremath{\pion^+}}\xspace}
\def\pim    {{\ensuremath{\pion^-}}\xspace}

\def\kaon    {{\ensuremath{\PK}}\xspace}
  \def\Kbar    {{\kern 0.2em\overline{\kern -0.2em \PK}{}}\xspace}

\def\KorKbar    {\kern 0.18em\optbar{\kern -0.18em K}{}\xspace}

\def\Kp      {{\ensuremath{\kaon^+}}\xspace}
\def\Km      {{\ensuremath{\kaon^-}}\xspace}

\def\KS      {{\ensuremath{\kaon^0_{\mathrm{ \scriptscriptstyle S}}}}\xspace}

  \def\Dbar    {{\kern 0.2em\overline{\kern -0.2em \PD}{}}\xspace}

\def\DorDbar    {\kern 0.18em\optbar{\kern -0.18em D}{}\xspace}

\def\B       {{\ensuremath{\PB}}\xspace}
\def\Bbar    {{\ensuremath{\kern 0.18em\overline{\kern -0.18em \PB}{}}}\xspace}
\def\Bb      {{\ensuremath{\Bbar}}\xspace}
\def\BorBbar    {\kern 0.18em\optbar{\kern -0.18em B}{}\xspace}
\def\Bz      {{\ensuremath{\B^0}}\xspace}

\def\Bd      {{\ensuremath{\B^0}}\xspace}
\def\Bs      {{\ensuremath{\B^0_\squark}}\xspace}
\def\Bsb     {{\ensuremath{\Bbar{}^0_\squark}}\xspace}
\def\Bdb     {{\ensuremath{\Bbar{}^0}}\xspace}

\def\jpsi     {{\ensuremath{{\PJ\mskip -3mu/\mskip -2mu\Ppsi\mskip 2mu}}}\xspace}

  \def\Y#1S{\ensuremath{\PUpsilon{(#1S)}}\xspace}

\def\Lz          {{\ensuremath{\PLambda}}\xspace}
\def\Lbar        {{\ensuremath{\kern 0.1em\overline{\kern -0.1em\PLambda}}}\xspace}
\def\LorLbar    {\kern 0.18em\optbar{\kern -0.18em \PLambda}{}\xspace}

\def\Lb      {{\ensuremath{\Lz^0_\bquark}}\xspace}

\def\to                 {\ensuremath{\rightarrow}\xspace}

\def\order   {{\ensuremath{\mathcal{O}}}\xspace}

\def\CP                {{\ensuremath{C\!P}}\xspace}
\def\CPT               {{\ensuremath{C\!PT}}\xspace}

\def\Vcd  {{\ensuremath{V_{\cquark\dquark}}}\xspace}
\def\Vtd  {{\ensuremath{V_{\tquark\dquark}}}\xspace}

\def\Vcs  {{\ensuremath{V_{\cquark\squark}}}\xspace}
\def\Vts  {{\ensuremath{V_{\tquark\squark}}}\xspace}

\def\Vcbs  {{\ensuremath{V_{\cquark\bquark}^\ast}}\xspace}
\def\Vtbs  {{\ensuremath{V_{\tquark\bquark}^\ast}}\xspace}

\newcommand{\dm}{{\ensuremath{\Delta m}}\xspace}
\newcommand{\dms}{{\ensuremath{\Delta m_{\squark}}}\xspace}
\newcommand{\dmd}{{\ensuremath{\Delta m_{\dquark}}}\xspace}
\newcommand{\DG}{{\ensuremath{\Delta\Gamma}}\xspace}
\newcommand{\DGs}{{\ensuremath{\Delta\Gamma_{\squark}}}\xspace}
\newcommand{\DGd}{{\ensuremath{\Delta\Gamma_{\dquark}}}\xspace}

\newcommand{\phis}{{\ensuremath{\phi_{\squark}}}\xspace}

\def\AT#1     {\ensuremath{A_{\mathrm{T}}^{#1}}\xspace}           

\def\C#1      {\ensuremath{\mathcal{C}_{#1}}\xspace}                       
\def\Cp#1     {\ensuremath{\mathcal{C}_{#1}^{'}}\xspace}                    
\def\Ceff#1   {\ensuremath{\mathcal{C}_{#1}^{\mathrm{(eff)}}}\xspace}        
\def\Cpeff#1  {\ensuremath{\mathcal{C}_{#1}^{'\mathrm{(eff)}}}\xspace}       
\def\Ope#1    {\ensuremath{\mathcal{O}_{#1}}\xspace}                       
\def\Opep#1   {\ensuremath{\mathcal{O}_{#1}^{'}}\xspace}                    

\newcommand{\ket}[1]{\ensuremath{|#1\rangle}}              

\newcommand{\tev}{\ifthenelse{\boolean{inbibliography}}{\ensuremath{~T\kern -0.05em eV}\xspace}{\ensuremath{\mathrm{\,Te\kern -0.1em V}}}\xspace}
\newcommand{\gev}{\ensuremath{\mathrm{\,Ge\kern -0.1em V}}\xspace}
\newcommand{\mev}{\ensuremath{\mathrm{\,Me\kern -0.1em V}}\xspace}
\newcommand{\kev}{\ensuremath{\mathrm{\,ke\kern -0.1em V}}\xspace}
\newcommand{\ev}{\ensuremath{\mathrm{\,e\kern -0.1em V}}\xspace}
\newcommand{\gevc}{\ensuremath{{\mathrm{\,Ge\kern -0.1em V\!/}c}}\xspace}
\newcommand{\mevc}{\ensuremath{{\mathrm{\,Me\kern -0.1em V\!/}c}}\xspace}
\newcommand{\gevcc}{\ensuremath{{\mathrm{\,Ge\kern -0.1em V\!/}c^2}}\xspace}
\newcommand{\gevgevcccc}{\ensuremath{{\mathrm{\,Ge\kern -0.1em V^2\!/}c^4}}\xspace}
\newcommand{\mevcc}{\ensuremath{{\mathrm{\,Me\kern -0.1em V\!/}c^2}}\xspace}

\def\invfb   {\ensuremath{\mbox{\,fb}^{-1}}\xspace}

\def\invps{\ensuremath{{\mathrm{ \,ps^{-1}}}}\xspace}

\def\hr   {\ensuremath{\mathrm{ \,hr}}\xspace}

\newcommand{\stat}{\ensuremath{\mathrm{\,(stat)}}\xspace}
\newcommand{\syst}{\ensuremath{\mathrm{\,(syst)}}\xspace}

\def\order{{\ensuremath{\mathcal{O}}}\xspace}

\def\deriv {\ensuremath{\mathrm{d}}}

\def\gsim{{~\raise.15em\hbox{$>$}\kern-.85em
          \lower.35em\hbox{$\sim$}~}\xspace}
\def\lsim{{~\raise.15em\hbox{$<$}\kern-.85em
          \lower.35em\hbox{$\sim$}~}\xspace}

\newcommand{\mean}[1]{\ensuremath{\left\langle #1 \right\rangle}} 
\newcommand{\Real}{\ensuremath{\mathcal{R}e}\xspace}
\newcommand{\Imag}{\ensuremath{\mathcal{I}m}\xspace}

\def\sPlot{\mbox{\em sPlot}\xspace}

\def\degrees{\ensuremath{^{\circ}}\xspace}

\def\mrad{\ensuremath{\mathrm{ \,mrad}}\xspace}

\def\tell1  {TELL1\xspace}
\def\ukl1   {UKL1\xspace}

\newcommand{\ie}{\mbox{\itshape i.e.}\xspace}

\usepackage{cite} 
\usepackage{mciteplus}

\def\myTitle {Search for violations of Lorentz invariance and $\CPT$ symmetry in  $\Bds$ mixing}

\def\DatVal  {-0.10}
\def\DatStat { 0.82}
\def\DatSyst { 0.54}
\def\DasVal  {-0.20}
\def\DasStat { 0.22}
\def\DasSyst { 0.04}
\def\DaxVal  {+1.97}
\def\DaxStat { 1.30}
\def\DaxSyst { 0.29}
\def\DayVal  {+0.44}
\def\DayStat { 1.26}
\def\DaySyst { 0.29}
\def\DamuUnit {\ensuremath{\times 10^{-15}\gev}}
\def\DasUnit  {\ensuremath{\times 10^{-13}\gev}}

\def\DatBsVal  {-0.89}
\def\DatBsStat { 1.41}
\def\DatBsSyst { 0.36}
\def\DasBsVal  {-0.47}
\def\DasBsStat { 0.39}
\def\DasBsSyst { 0.08}
\def\DaxBsVal  {+1.01}
\def\DaxBsStat { 2.08}
\def\DaxBsSyst { 0.71}
\def\DayBsVal  {-3.83}
\def\DayBsStat { 2.09}
\def\DayBsSyst { 0.71}
\def\DamuBsUnit {\ensuremath{\times 10^{-14}\gev}}
\def\DasBsUnit  {\ensuremath{\times 10^{-12}\gev}}

\def\RezVal  {-0.022}
\def\RezStat { 0.033}
\def\RezSyst { 0.003}
\def\ImzVal  {\phantom{+}0.004}
\def\ImzStat { 0.011}
\def\ImzSyst { 0.002}

\def\Bds     {{\ensuremath{\B^0_{\kern -0.1em{\scriptscriptstyle (}\kern -0.05em\squark\kern -0.03em{\scriptscriptstyle )}}}}\xspace}

\def\Dan          {{\ensuremath{\Delta a_0}}\xspace}

\def\Dax          {{\ensuremath{\Delta a_X}}\xspace}
\def\DaxBd        {{\ensuremath{\Delta a_X^{\Bd}}}\xspace}
\def\DaxBs        {{\ensuremath{\Delta a_X^{\Bs}}}\xspace}
\def\Day          {{\ensuremath{\Delta a_Y}}\xspace}
\def\DayBd        {{\ensuremath{\Delta a_Y^{\Bd}}}\xspace}
\def\DayBs        {{\ensuremath{\Delta a_Y^{\Bs}}}\xspace}
\def\Daz          {{\ensuremath{\Delta a_Z}}\xspace}

\def\Daxy         {{\ensuremath{\Delta a_{X,Y}}}\xspace}
\def\Damu         {{\ensuremath{\Delta a_{\mu}}}\xspace}

\def\DamuBd       {{\ensuremath{\Delta a^{\Bd}_{\mu}}}\xspace}
\def\DamuBs       {{\ensuremath{\Delta a^{\Bs}_{\mu}}}\xspace}
\def\Dat          {{\ensuremath{\Dan - 0.38 \Daz}}\xspace}

\def\Das          {{\ensuremath{0.38 \Dan + \Daz}}\xspace}

\def\Dap          {{\ensuremath{\Delta a_{\parallel}}}\xspace}
\def\DapBd        {{\ensuremath{\Delta a_{\parallel}^{\Bd}}}\xspace}
\def\DapBs        {{\ensuremath{\Delta a_{\parallel}^{\Bs}}}\xspace}
\def\Dao          {{\ensuremath{\Delta a_{\perp}}}\xspace}
\def\DaoBd        {{\ensuremath{\Delta a_{\perp}^{\Bd}}}\xspace}
\def\DaoBs        {{\ensuremath{\Delta a_{\perp}^{\Bs}}}\xspace}
\def\Daxy         {{\ensuremath{\Delta a_{X,Y}}}\xspace}

\newcommand{\Ap}{\ensuremath{A_P}\xspace}

\def\qf           {{\ensuremath{\zeta}}\xspace}

\def\Re           {{\ensuremath{\Real}}\xspace}
\def\Im           {{\ensuremath{\Imag}}\xspace}

\def\Rez           {{\ensuremath{\Re(z)}}\xspace}
\def\Imz           {{\ensuremath{\Im(z)}}\xspace}
\def\RezBs         {{\ensuremath{\Re(z^{\Bs})}}\xspace}
\def\ImzBs         {{\ensuremath{\Im(z^{\Bs})}}\xspace}
\def\RezBd         {{\ensuremath{\Re(z^{\Bd})}}\xspace}

\def\zBs           {{\ensuremath{z^{\Bs}}}\xspace}

\def\Rijp         {{\ensuremath{\eta^+}}\xspace}
\def\Rijm         {{\ensuremath{\eta^-}}\xspace}
\def\Rijpm        {{\ensuremath{\eta^{\pm}}}\xspace}
\def\RijR         {{\ensuremath{\eta^{\Re}}}\xspace}
\def\RijI         {{\ensuremath{\eta^{\Im}}}\xspace}

\def\Cf           {{\ensuremath{C_f}}\xspace}
\def\Sf           {{\ensuremath{S_f}}\xspace}
\def\Df           {{\ensuremath{D_f}}\xspace}

\def\AS           {{\ensuremath{A_{\mathrm S}}}\xspace}
\def\Az           {{\ensuremath{A_0}}\xspace}
\def\Ap           {{\ensuremath{A_{\parallel}}}\xspace}
\def\Ao           {{\ensuremath{A_{\perp}}}\xspace}

\def\Aj           {{\ensuremath{A_m}}\xspace}

\def\ASs          {{\ensuremath{A_{\mathrm S}^*}}\xspace}
\def\Azs          {{\ensuremath{A_0^*}}\xspace}
\def\Aps          {{\ensuremath{A_{\parallel}^*}}\xspace}

\def\Ais          {{\ensuremath{A_l^*}}\xspace}

\DeclareGraphicsExtensions{.pdf,.PDF,png,.PNG}

\newboolean{wordcount}
\setboolean{wordcount}{false} 

\def\figWidth{0.75}
\def\breakEquation{}
\def\breakEqLine{}
\def\breakEqColumn{}

\begin{document}

\renewcommand{\thefootnote}{\fnsymbol{footnote}}
\setcounter{footnote}{1}

\begin{titlepage}
\pagenumbering{roman}

\vspace*{-1.5cm}
\centerline{\large EUROPEAN ORGANIZATION FOR NUCLEAR RESEARCH (CERN)}
\vspace*{1.5cm}
\noindent
\begin{tabular*}{\linewidth}{lc@{\extracolsep{\fill}}r@{\extracolsep{0pt}}}
\ifthenelse{\boolean{pdflatex}}
{\vspace*{-2.7cm}\mbox{\!\!\!\includegraphics[width=.14\textwidth]{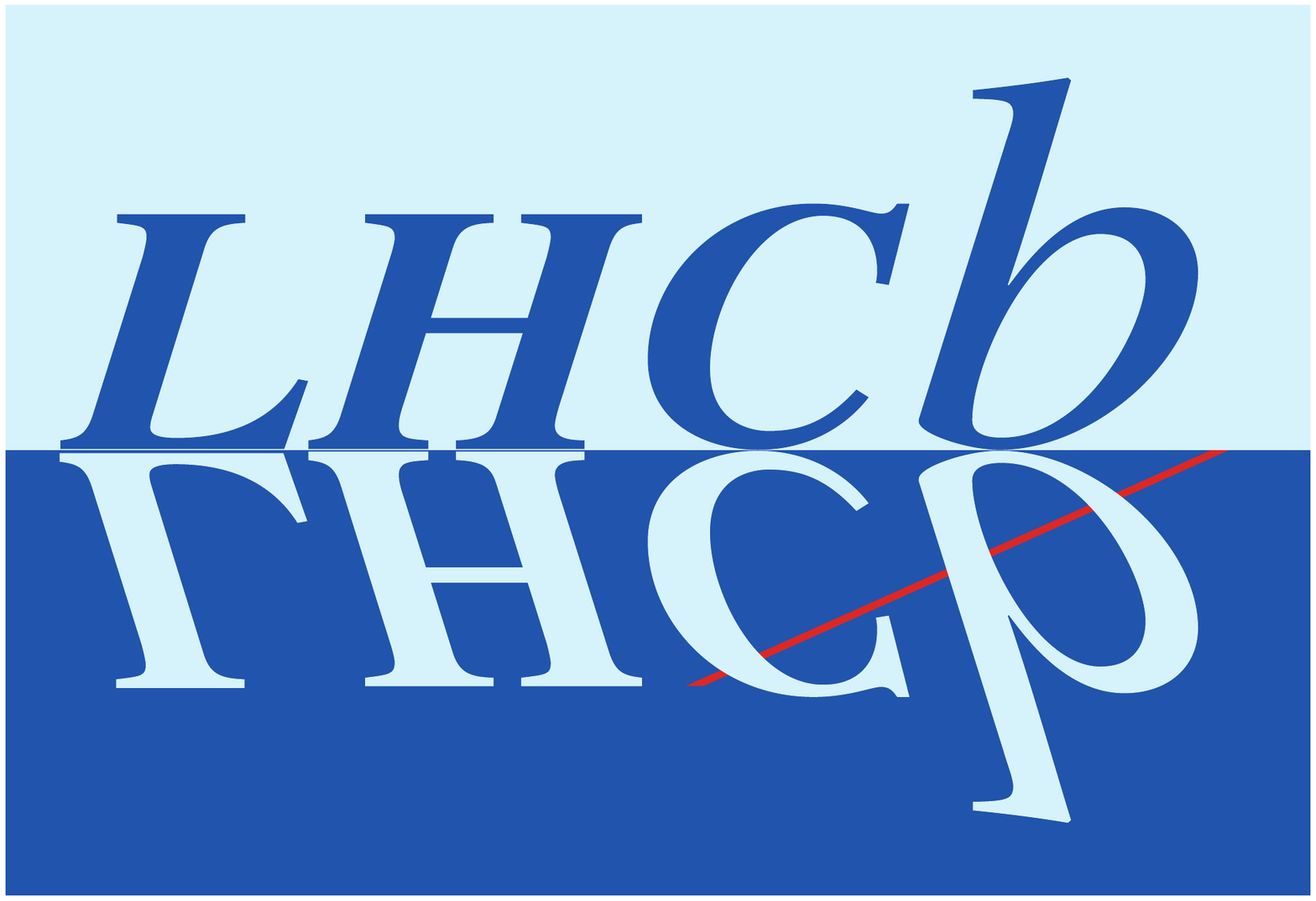}} & &}%
{\vspace*{-1.2cm}\mbox{\!\!\!\includegraphics[width=.12\textwidth]{lhcb-logo.eps}} & &}%
\\
 & & CERN-EP-2016-048 \\  
 & & LHCb-PAPER-2016-005 \\  
 & & June 16, 2016 \\ 
\end{tabular*}

\vspace*{2.0cm}

{\normalfont\bfseries\boldmath\huge
\begin{center}
  \myTitle
\end{center}
}

\vspace*{1.0cm}

\begin{center}
The LHCb collaboration\footnote{Authors are listed at the end of this Letter.}
\end{center}

\vspace{\fill}

\begin{abstract}
  \noindent
  Violations of $\CPT$ symmetry and Lorentz invariance are searched for by
studying interference effects in $\Bd$ mixing and in $\Bs$ mixing. Samples of
$\Bd\to\jpsi\KS$ and $\Bs\to\jpsi\Kp\Km$ decays are recorded by the LHCb
detector in proton--proton collisions at centre-of-mass energies of 7 and 8\tev,
corresponding to an integrated luminosity of 3\invfb. No periodic variations of
the particle-antiparticle mass differences are found, consistent with Lorentz
invariance and $\CPT$ symmetry. Results are expressed in terms of the Standard
Model Extension parameter $\Delta a_{\mu}$ with precisions of $\order(10^{-15})$
and $\order(10^{-14})$\gev for the $\Bd$ and $\Bs$ systems, respectively. With
no assumption on Lorentz (non-)invariance, the $\CPT$-violating parameter $z$ in
the $\Bs$ system is measured for the first time and found to be $\Rez = -0.022
\pm 0.033 \pm 0.005$ and $\Imz = 0.004 \pm 0.011\pm 0.002$, where the first
uncertainties are statistical and the second systematic.

\end{abstract}

\vspace*{1.0cm}

\begin{center}
    Published in Phys.~Rev.~Lett. 116 (2016) 241601 
\end{center}

\vspace{\fill}

{\footnotesize 
\centerline{\copyright~CERN on behalf of the \lhcb collaboration, licence \href{http://creativecommons.org/licenses/by/4.0/}{CC-BY-4.0}.}}
\vspace*{2mm}

\end{titlepage}


\newpage
\setcounter{page}{2}
\mbox{~}

\cleardoublepage


\renewcommand{\thefootnote}{\arabic{footnote}}
\setcounter{footnote}{0}


\pagestyle{plain} 
\setcounter{page}{1}
\pagenumbering{arabic}


\noindent
Lorentz invariance and the combination of charge conjugation, spatial inversion
and time reversal (\CPT) are exact symmetries in the Standard Model (SM) of
particle physics, and are deeply connected in any quantum field
theory~\cite{Greenberg:2002uu}. Quantum theories that aim to describe
Planck-scale physics, such as string theory, might break these fundamental
symmetries~\cite{PhysRevD.39.683}. Present-day experiments are many orders of
magnitude away from the Planck energy scale of $\sim10^{19}\gev$, however, small
effects at low energy might still be observable. Interference effects in the
mixing of neutral mesons are sensitive to violations of \CPT symmetry, and
therefore may provide a window to the quantum gravity
scale~\cite{Liberati:2013xla}. Such effects can be quantified in a low-energy,
effective field theory, as done in the Standard Model Extension
(SME)~\cite{Colladay:1996iz,Colladay:1998fq}. In this framework, terms that
explicitly break Lorentz and \CPT symmetry are added to the SM Lagrangian to
describe the couplings between particles and (hypothetical) uniform tensor
fields. These fields would acquire non-zero vacuum expectation values when
these symmetries are spontaneously broken in the underlying theory. The SME
couplings are expected to be suppressed by powers of the Planck
scale~\cite{Kostelecky:1991ak,*Kostelecky:1994rn}. In the SME, the
\CPT-violating parameters that can be measured in neutral meson systems also
break Lorentz symmetry. The amount of \CPT violation depends on the direction
of motion and on the boost of the particle. The SME parameters for the \Bd and
\Bs systems can be best measured with a time-dependent analysis of
the decay channels $\Bz\to\jpsi\KS$ and $\Bs\to\jpsi\Kp\Km$, using the
four-velocity of the \B mesons~\cite{vanTilburg:2014dka}. The notation \B refers
to either \Bd or \Bs and the inclusion of charge-conjugate processes is implied
throughout this Letter. These parameters have been measured previously, albeit
with less sensitive decay modes~\cite{vanTilburg:2014dka}, by the
\babar collaboration for the \Bd system~\cite{Aubert:2007bp}, and by the \dzero
collaboration for the \Bs system~\cite{Abazov:2015ana}.

The LHCb detector is a single-arm forward spectrometer described in detail in
Refs.~\cite{Alves:2008zz,LHCb-DP-2014-002}. Simulated events are produced using
the software described in Refs.~\cite{Sjostrand:2006za,*Sjostrand:2007gs,
  LHCb-PROC-2010-056, Lange:2001uf, Allison:2006ve, *Agostinelli:2002hh,
  LHCb-PROC-2011-006}.  The data used in this analysis correspond to an
integrated luminosity of $3\invfb$, taken at the LHC at proton--proton
centre-of-mass energies of 7 and 8\tev. The selection of both decay channels is
the same as used in Refs.~\cite{LHCb-PAPER-2015-004} and
\cite{LHCb-PAPER-2014-059}. The \jpsi meson is reconstructed in the dimuon
channel and the \KS meson in the \pip\pim final state.

Interference effects from \CPT violation can be incorporated generically in the
time evolution of a neutral \B meson system, described by the Schr\"{o}dinger
equation $i \partial_t \Psi = \hat{H} \Psi$. The effective $2\times2$
Hamiltonian is written as $\hat{H}=\hat{M}-i\hat{\Gamma}/2$~\cite{PDG2014}.
Diagonalisation gives a heavy-mass eigenstate $\ket{B_{H}}$ and a light-mass
eigenstate $\ket{B_{L}}$ with masses $m_{H,L}$ and decay widths
$\Gamma_{H,L}$. The differences between the eigenvalues 
are defined as $\dm \equiv m_{H} - m_{L}$ and $\DG \equiv
\Gamma_{L} - \Gamma_{H}$. The differences between the diagonal
matrix elements of the effective Hamiltonian are defined as 
$\delta m \equiv M_{11} - M_{22}$ and $\delta \Gamma
\equiv \Gamma_{11} - \Gamma_{22}$. Any difference between the mass or
lifetime of particles and antiparticles (\ie, a non-zero $\delta m$ or $\delta
\Gamma$) would be a sign of \CPT violation, and is characterised by
\begin{align}
  z = \frac{\delta m - i \delta \Gamma/2}{\dm + i\DG/2} \ ,
\label{eq:zdef}
\end{align}
and the mass eigenstates are given by $\ket{B_{H,L}} = p \sqrt{1\pm z} \ket{\B}
\mp q \sqrt{1\mp z} \ket{\Bb}$. Owing to the smallness of the \B mixing
parameters $\dm$ and $\DG$ in the denominator, $z$ is highly sensitive to
\CPT-violating effects.

Considering only contributions to first order in $z$, the decay rate to a \CP
eigenstate $f$ as a function of the \B proper decay time $t$ becomes
\begin{align}
    \frac{\deriv \Gamma_f}{\deriv t}
    \propto e^{-\Gamma t} &
    \bigg[ \big[1 + \qf \Df \Rez - \Sf \Imz\big] \cosh(\DG t/2) \nonumber \\ &
      + \big[\Df +  \Rez(\Cf+\qf)\big]  \sinh(\DG t/2) \nonumber \\ 
      & +\qf \big[ \Cf - \Df \Rez +\qf \Sf \Imz \big] \cos(\dm t)\nonumber \\ & 
      -\qf \big[ \Sf - \Imz(\Cf+\qf) \big] \sin(\dm t)  \bigg] \ , 
\label{eq:mastercp}
\end{align} 
where $\Gamma\equiv(\Gamma_{11}+\Gamma_{22})/2$, $\qf=+1(-1)$ for an initial
\ket{\B} (\ket{\Bb}) state and the following definitions are introduced:
\begin{align}
  \Cf \equiv \frac{1-|\lambda_f|^2}{1+|\lambda_f|^2} \ , \ \breakEqColumn
  \Sf \equiv \frac{2\Im(\lambda_f)}{1+|\lambda_f|^2} \ , \ \breakEqLine
  \Df \equiv -\frac{2\Re(\lambda_f)}{1+|\lambda_f|^2} \ , \ \breakEqColumn
  \lambda_f \equiv \frac{q}{p}\frac{\overline{A}_f}{A_f} \ ,
  \label{eq:CSD}
\end{align}
with $A_f$ and $\overline{A}_f$ the direct decay amplitudes of a \ket{\B} or
\ket{\Bb} state to the eigenstate $f$.

For the decay $\Bz\to\jpsi\KS$, the final state is \CP odd, corresponding to \CP
eigenvalue $\eta_f=-1$. In the SM, $\arg(\lambda_{\jpsi\KS})=\pi-2\beta$, where
$\beta$ is defined in terms of elements of the CKM matrix as
$\beta \equiv \arg[-(\Vcd\Vcbs)/(\Vtd\Vtbs)]$. Furthermore, in the \Bd system,
the approximation $\DGd=0$ is made, as supported by experimental
data~\cite{HFAG}.

The decay $\Bs\to\jpsi\Kp\Km$ is similar to $\Bz\to\jpsi\KS$, but the decay
width difference \DGs cannot be ignored~\cite{HFAG}. Another important
difference is that the $\Kp\Km$ system mostly originates from the $\phi(1020)$
resonance, giving the $\Kp\Km$ pair an orbital angular momentum $L=1$ (P
wave). Since the $\jpsi\phi$ final state consists of two vector mesons, its
orbital angular momentum can be $L \in \{0,1,2\}$ for the polarisation states $f
\in \{0,\perp,\parallel\}$, respectively, with corresponding \CP eigenvalues
$\eta_f=(-1)^L$. The $\Kp\Km$ system has a small S-wave
contribution~\cite{LHCb-PAPER-2014-059}, which results in another $L=1$
component for the $\jpsi\Kp\Km$ final state. These four polarisation states can
be separated statistically in the helicity formalism~\cite{LHCb-PAPER-2013-002},
using the three decay angles between the final-state particles. The
corresponding weak phases, $\arg(\lambda_{\jpsi\Kp\Km}) = L\pi-\phis$, can, in
the SM, be expressed in terms of CKM matrix elements,
$\phis=-2\beta_s\equiv-2\arg[-(\Vts\Vtbs)/(\Vcs\Vcbs)]$.  The decay rate has to
be modified compared to Eq.~\ref{eq:mastercp} to include the angular
dependence. It becomes a sum over all ten combinations of the four helicity
amplitudes,
\begin{align}
  \frac{\deriv^{4}\Gamma_{\jpsi\Kp\Km}}{\deriv t\,\deriv\vec{\Omega}} 
  \propto \sum_{k=1}^{10}h_k(t)f_k(\vec{\Omega}) \ ,
\label{eq:BsSum}
\end{align}
where $f_k(\vec{\Omega})$ are angular functions, given in
Ref.~\cite{LHCb-PAPER-2013-002}, and $h_k(t)$ are products of the amplitudes as
listed in Table~\ref{tab:BsTimeDep}. The time dependence of $h_k(t)$ is given
by
\begin{align}
  \Ais(t)\Aj(t) = &\frac{\Ais(0) \Aj(0) e^{-\Gamma_s t}}{1+\qf \Cf} \nonumber \\
  & \bigg[ a_k \cosh(\DGs t/2) + b_k \sinh(\DGs t/2) \nonumber \\
  & \;\; + c_k \cos(\dms t) + d_k \sin(\dms t) \bigg] \ ,
\label{eq:BsAmp}
\end{align} 
with the coefficients listed in Table~\ref{tab:BsCoef}.

\begin{table}
  \centering
  \caption{Time-dependent functions $h_k(t)$ in Eq.~\ref{eq:BsSum}.}
    \label{tab:BsTimeDep}
    \ifthenelse{\boolean{wordcount}}{}{
    \begin{tabular}{cc}\hline
      $k$  & $h_k(t)$             \\ \hline
      $1$  & $|\Az(t)|^2$         \\
      $2$  & $|\Ap(t)|^2$         \\
      $3$  & $|\Ao(t)|^2$         \\
      $4$  & $\Im(\Aps(t)\Ao(t))$ \\
      $5$  & $\Re(\Azs(t)\Ap(t))$ \\ \hline
    \end{tabular}
    \quad
    \begin{tabular}{cc}\hline
      $k$  & $h_k(t)$ \\ \hline
      $6$  & $\Im(\Azs(t)\Ao(t))$ \\
      $7$  & $|\AS(t)|^2$         \\
      $8$  & $\Re(\ASs(t)\Ap(t))$ \\
      $9$  & $\Im(\ASs(t)\Ao(t))$ \\ 
      $10$ & $\Re(\ASs(t)\Az(t))$ \\ \hline
  \end{tabular}}
\end{table}

\begin{table}
\centering
\caption{Definition of the coefficients in Eq.~\ref{eq:BsAmp}. The 
following definitions are used:
$\Rijp \equiv (1+\eta_{l}\eta_{m})/2$, 
$\Rijm \equiv (1-\eta_{l}\eta_{m})/2$,
$\RijI \equiv i(\eta_{l}-\eta_{m})/2$,
$\RijR \equiv  (\eta_{l}+\eta_{m})/2$. Furthermore, 
$\qf^{+} \equiv (\qf)^{\Rijp}$, and
$\qf^{-} \equiv (\qf)^{\Rijm}$, such that $\qf^{\pm}=1$ if $\Rijpm=0$ and $\qf^{\pm}=\qf$ otherwise.
}
\label{tab:BsCoef}
\ifthenelse{\boolean{wordcount}}{}{
\begin{tabular}{cl} \hline
  $a_k=$ & $(\Rijp + \Rijm \Cf) + \qf \Rez(\RijR \Df + \RijI \Sf)$ \\
         \multicolumn{2}{r}{ $+ \Imz (\RijI \Df - \RijR \Sf)\;\;$} \\
  $b_k=$ & $(\RijR \Df + \RijI \Sf) + \Rez (\qf^{+} + \qf^{-} \Cf)$ \\
  $c_k$= & $\qf (\Rijm + \Rijp \Cf) - \qf \Rez(\RijR \Df + \RijI \Sf)$ \\
         \multicolumn{2}{r}{$- \Imz (\RijI \Df - \RijR \Sf)$} \\
  $d_k$= & $\qf(\RijI \Df - \RijR \Sf) + \Imz  (\qf^{+} + \qf^{-} \Cf) $\\
\hline
\end{tabular}}
\end{table}

In the SME, the dimensionless parameter $z$ is not a constant. It depends on the
four-velocity $\beta^{\mu}=(\gamma,\gamma\vec{\beta})$ of the neutral meson
as~\cite{Kostelecky:1997mh, Kostelecky:2001ff}
\begin{align}
z = \frac{\beta^{\mu}\Damu}{\dm + i\DG/2} \ ,
\label{eq:zSME}
\end{align}
thereby breaking Lorentz invariance. The SME parameter \Damu describes the
difference between the couplings of the valence quarks, within the neutral
meson, with the Lorentz-violating fields~\cite{Kostelecky:1997mh}. Therefore,
\Bd and \Bs mesons can have different values of \Damu.  Since \Damu is
real~\cite{Kostelecky:1999bm}, it follows that $\Rez \DG = - 2 \Imz \dm$. For \B
mesons, $\dm\gg\DG$, and so \Imz is two orders of magnitude smaller than \Rez,
and can be ignored in the measurements of \Damu. The average boost of \B mesons
in the acceptance of \lhcb is $\mean{\gamma\beta} \approx 20$. It follows from
Eq.~\ref{eq:zSME} that this large boost results in a high sensitivity to
\Damu~\cite{vanTilburg:2014dka}.

To measure \Damu, the meson direction needs to be determined in an absolute
reference frame. Such a frame can be defined with respect to fixed
stars~\cite{Kostelecky:1999bm}. In this frame, the $Z$-axis points north along
the Earth's rotation axis, the $X$-axis points from the Sun to the vernal
equinox on 1 January 2000 (J2000 epoch) and the $Y$-axis completes the
right-handed coordinate system. The latitude of the \lhcb interaction point is
$46.2414\degrees$N, the longitude is $6.0963\degrees$E, and the angle of the
beam east of north is $236.296\degrees$. The beam axis is inclined with respect
to the geodetic plane by $3.601\mrad$, pointing slightly upwards.  The
timekeeping is obtained from the LHC machine with a time stamp, $t_{\rm LHC}$,
in UTC microseconds since January $1^{\rm st}$, 1970, 00:00:00 UTC. The time,
spatial coordinates and angles have negligible uncertainties and are used to
define the rotation from the coordinate system of \lhcb to the absolute
reference frame. For mesons travelling along the beam axis, \Rez can be
expressed as,
\begin{align}
  \Rez & = \frac{\dm}{\dm^2+\DG^2/4} \beta^{\mu}\Damu  \nonumber \\
  & \approx \frac{\gamma}{\dm} 
  \Big[ \Dan + \cos(\chi) \Daz \breakEquation 
    +\sin(\chi) \big[ \Day \sin(\Omega \hat{t}) + \Dax
      \cos(\Omega \hat{t}) \big]\Big] \ ,
\label{eq:siddamu}
\end{align}
where $|\vec{\beta}|$ is set to unity, $\Delta a^{X,Y,Z} = -\Delta a_{X,Y,Z}$, and
$\chi=112.4\degrees$ is the angle between the beam axis and the rotational axis
of the Earth. The time dependence results from the Earth's rotation, giving a
periodicity with sidereal frequency $\Omega$. The sidereal phase at $t_{\rm
  LHC}=0$ is found to be $\hat{t} = (2.8126 \pm 0.0014)\hr$. The \B mesons are
emitted at an average angle of about $5\degrees$ from the beam axis. This means
that the \lhcb detector is mostly sensitive to the linear combination $\Dap
\equiv \Dan+\cos(\chi)\Daz= \Dat$, while there is a much weaker sensitivity to
the orthogonal parameter, $\Dao=\Das$, coming from the smaller transverse
component of the \B velocity. Both \Dap and \Dao are measured and the
correlation between them is negligible.

Unbinned likelihood fits are applied to the decay-time distributions using
Eqs.~\ref{eq:mastercp} and \ref{eq:BsSum}. To obtain the SME parameters, the
sidereal variation of \Rez is taken into account by including in the fits the
LHC time and the three-momentum of the reconstructed \B candidate. For the \Bs
sample, the fits are performed to the full angular distribution.

In the invariant mass distributions of the \B candidates, the background is
mostly combinatorial. For both decay channels, this background is statistically
subtracted using the \sPlot technique~\cite{Pivk:2004ty},
which allows to project out the signal component
by weighting each event depending on the mass of the \B candidate.
The mass models are the
same as in Refs.~\cite{LHCb-PAPER-2015-004, LHCb-PAPER-2014-059}. The
correlation between the shape of the invariant mass distribution and the \B
momentum or, for the \Bs sample, the decay angles, leads to a small systematic
bias for both samples. This effect is included in the systematic uncertainty.
In the \Bs\to\jpsi\Kp\Km sample, there is a small contribution coming from
misidentified \Bd\to\jpsi\Kp\pim and $\Lb\to\jpsi\Pp\Km$ decays. This background
contribution is statistically removed by adding simulated decays with negative
weights. A systematic uncertainty is assigned to account for the uncertainty on
the size and shape of this background.

The description of the detection efficiency as a function of the decay time, the
decay-time resolution model and the flavour tagging (to distinguish initial \B
and \Bbar mesons) are the same as in Refs.~\cite{LHCb-PAPER-2015-004} and
\cite{LHCb-PAPER-2014-059} for the $\Bd\to\jpsi\KS$ and $\Bs\to\jpsi\Kp\Km$
samples, respectively. This description includes the dilution of the asymmetry
due to wrong decisions of the flavour tagging method. The decay-time resolution
model and tagging calibration do not lead to a systematic bias in the final
result. A possible wrong assignment of the primary interaction vertex (PV) to
the \B candidate gives a small bias in the \DaoBd parameter, which is included
in the systematic uncertainty. The inefficiency at high decay times, caused by
the reconstruction algorithms, is described by an exponential function. For the
\Bd sample, this function is obtained from simulation and does not lead to a
systematic bias in the result. For the \Bs sample, the exponential function is
obtained from a data-driven method. The change in the final result when using
the correction procedure from Ref.~\cite{LHCb-PAPER-2014-059} is taken as a
systematic uncertainty.

The production asymmetry between \Bd and \Bdb mesons is included in the
modelling of the decay rates, and is taken from Refs.~\cite{LHCb-PAPER-2014-042,
  LHCb-PAPER-2014-053}. The corresponding uncertainties are included in the
statistical uncertainty, while a possible momentum dependence of the production
asymmetry is considered as a systematic uncertainty. The \Bs production
asymmetry does not affect the fit to the \Bs sample, since the fast \Bs
oscillations wash out this effect and since the decay rates for \Bs and \Bsb
tags are normalised separately.

In the fit to the \Bd sample, the correlation between \RezBd and $C_{\jpsi\KS}$
is too large to allow determination of \RezBd without making assumptions about
the value of $C_{\jpsi\KS}$~\cite{vanTilburg:2014dka}. On the other hand, to
determine \DamuBd, the averages $C_{\jpsi\KS}=0.005 \pm 0.020$ and
$S_{\jpsi\KS}=0.676 \pm 0.021$~\cite{PDG2014} as measured by the \babar and
\belle collaborations can be used in the fit. Since the boost of the \Bd mesons
is about 40 times lower in these experiments, these values are hardly affected
by possible Lorentz violation in the SME. The value of $D_{\jpsi\KS}$ is by
definition $\sqrt{\vphantom{S^2}}\overline{1 - S_{\jpsi\KS}^2 -
  C_{\jpsi\KS}^2}$. The uncertainties on these external input values are
propagated as systematic uncertainties on \DamuBd. The mass difference, $\dmd =
0.510 \pm 0.003\invps$~\cite{PDG2014}, is allowed to vary in the fit within its
uncertainty using a Gaussian constraint. Setting $\DGd=0.007\invps$, which
corresponds to the experimental uncertainty~\cite{HFAG}, leads to a small change
in \DapBd, which is included in the systematic uncertainty. The \Bd lifetime is
allowed to vary freely in the fit.

In the fit to the \Bs sample, the correlation between \RezBs and
$C_{\jpsi\Kp\Km}$ is small owing to the additional interference terms from the
helicity amplitudes, the non-zero \DGs and the faster \Bs--\Bsb
oscillations. For this reason, the same parameters as in
Ref.~\cite{LHCb-PAPER-2014-059} are varied freely in the fit, in addition to
either \Damu or $z$. The detection efficiency is also described as a function
of the decay angles. The shape of this angular acceptance is obtained from
simulation. The simulated events are weighted to match the kinematic
distributions in data. The uncertainty due to the limited number of simulated
events and the full effect of correcting for the kinematic distributions in data
are added to the systematic uncertainty. Systematic effects due to the
decay-angle resolution are negligibly small. The fit to the \Bs sample is
performed simultaneously in bins of the \Kp\Km invariant
mass~\cite{LHCb-PAPER-2014-059}. Each bin has a different interference between
the P- and S-wave amplitudes. This effect is included in the fit and no
systematic biases are observed.

An overview of the systematic uncertainties is given in
Table~\ref{tab:syst}. For the \Bd mixing, the largest contribution comes from
the uncertainty on the external parameters $C_f$ and $S_f$. A small systematic
bias is observed in \DaoBd due to the momentum dependence of the cross-sections
of neutral kaons in the detector material. For the \Bs mixing, the largest
contribution comes from the description of the decay-time acceptance. Effects
from the correlation between the mass and decay time and from the accuracy of
the length scale and momentum scale of the detector are found to be negligible.

\begin{table}
  \centering
  \caption{Systematic uncertainties on \Damu for \Bd mixing and on \Damu and $z$
    for \Bs mixing. Contributions marked with -- are negligible.}
  \label{tab:syst} 
\ifthenelse{\boolean{wordcount}}{}{
{\small \begin{center}
\begin{tabular}{lccc}
\hline
{\boldmath \Bd \bf mixing}   & \Dap & \Dao & \Daxy  \\            
Source                       & \multicolumn{3}{c}{[\DamuUnit]} \\ \hline     
Mass correlation             &   -- &  --  & 0.04 \\ 
Wrong PV assignment          &   -- &   1  &   -- \\
Production asymmetry         & 0.28 &   1  & 0.05 \\ 
External input $C_f$, $S_f$  & 0.46 &   4  & 0.28 \\
Decay width difference       & 0.07 &  --  &   -- \\ 
Neutral kaon asymmetry       &   -- &   1  &   -- \\  \hline  
Quadratic sum                & 0.54 &   4  & 0.29 \\ 
\hline
\end{tabular}
\end{center}}
{\small \begin{center}
\begin{tabular}{lccc|cc}
\hline
{\boldmath\Bs\bf mixing}& \Dap & \Dao & \Daxy & \Rez   & \Imz  \\
Source                  &\multicolumn{3}{c|}{[\DamuBsUnit]}& & \\ \hline
Mass correlation        & 0.10 &   3  & 0.24  & 0.001  & 0.002 \\
Peaking background      & 0.14 &   3  & 0.15  & 0.003  &    -- \\
Decay-time acceptance   & 0.30 &   7  & 0.65  &    --  & 0.001 \\
Angular acceptance      & 0.07 &  --  &   --  & 0.002  & 0.001 \\ \hline
Quadratic sum           & 0.36 &   8  & 0.71  & 0.003  & 0.002 \\ \hline  
\end{tabular}
\end{center}}}
\end{table}

\begin{figure}
  \centering
  \includegraphics[width=\figWidth\textwidth]{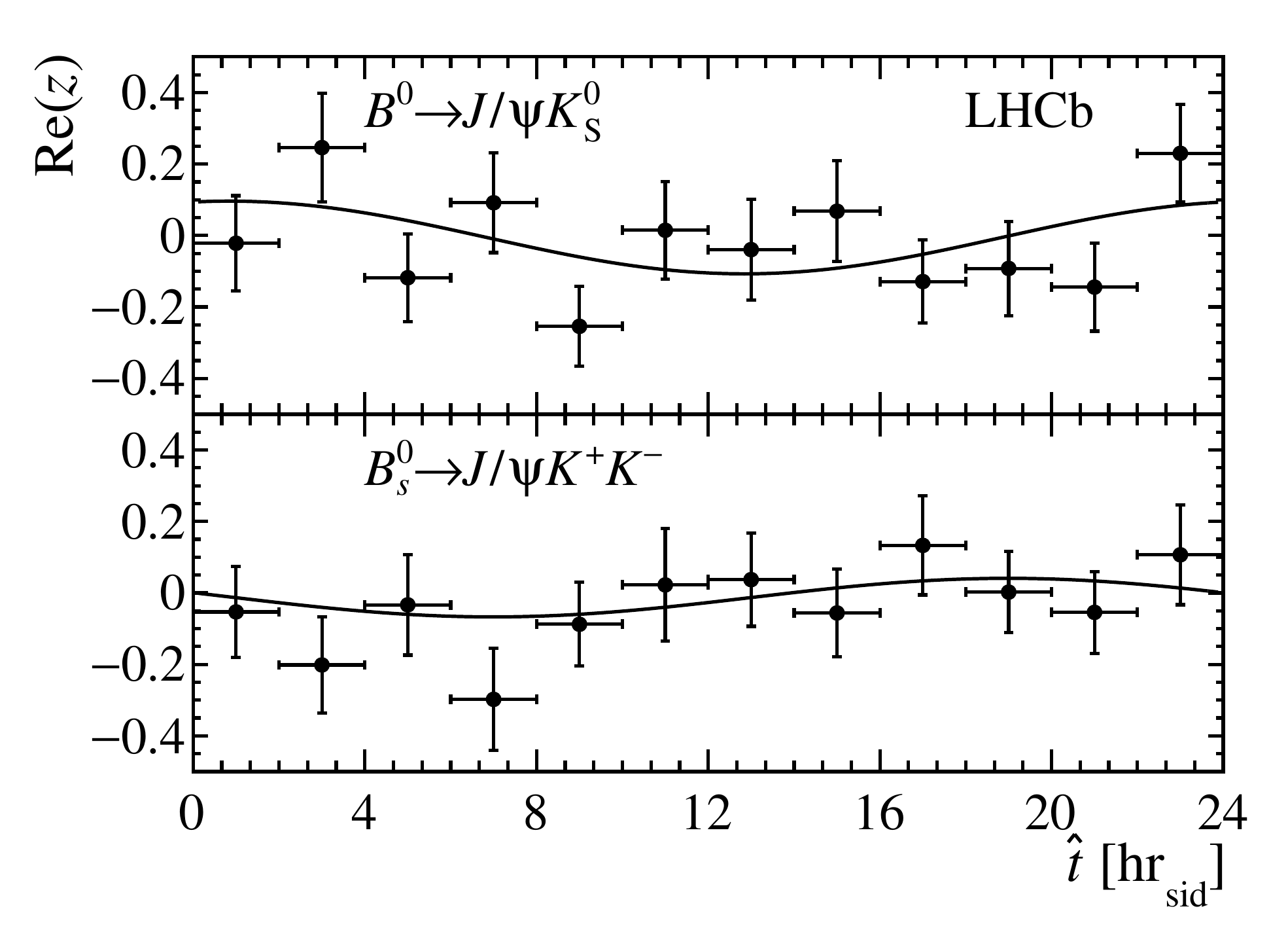}
  \caption{Values of \Rez obtained from fits in bins of sidereal phase for (top)
    the \Bd sample and (bottom) the \Bs sample. The solid line shows the
    variation of \Rez from the \Damu fits, using the average \B momentum.}
  \label{fig:rezplot}
\end{figure}

The components of the SME parameter \Damu for \Bd mixing, obtained from the
fit to the sample of selected \Bd\to\jpsi\KS candidates, are
\begin{align*}
  \DapBd &= (\DatVal \pm \DatStat\stat \pm \DatSyst\syst) \DamuUnit \ ,\\
  \DaoBd &= (\DasVal \pm \DasStat\stat \pm \DasSyst\syst) \DasUnit \ ,\\
  \DaxBd &= (\DaxVal \pm \DaxStat\stat \pm \DaxSyst\syst) \DamuUnit \ ,\\
  \DayBd &= (\DayVal \pm \DayStat\stat \pm \DaySyst\syst) \DamuUnit \ ,
\end{align*} 
and the corresponding numbers for \Bs mixing, using \Bs\to\jpsi\Kp\Km
candidates, are
\begin{align*}
  \DapBs &=(\DatBsVal \pm \DatBsStat\stat \pm \DatBsSyst\syst)\DamuBsUnit \ ,\\
  \DaoBs &=(\DasBsVal \pm \DasBsStat\stat \pm \DasBsSyst\syst)\DasBsUnit \ ,\\
  \DaxBs &=(\DaxBsVal \pm \DaxBsStat\stat \pm \DaxBsSyst\syst)\DamuBsUnit \ ,\\
  \DayBs &=(\DayBsVal \pm \DayBsStat\stat \pm \DayBsSyst\syst)\DamuBsUnit \ .
\end{align*} 
Figure~\ref{fig:rezplot} shows the result of fits of \Rez in bins of the
sidereal phase for both samples. For the \Bd sample, the external constraints
on $C_{\jpsi\KS}$ and $S_{\jpsi\KS}$ are again used. No sidereal
variation is observed. Independently of any assumption of Lorentz violation,
the complex \CPT-violating parameter $z$ in the \Bs system is found to be
\begin{align*}
  \RezBs = \RezVal \pm \RezStat\stat \pm \RezSyst\syst \ , \\
  \ImzBs = \ImzVal \pm \ImzStat\stat \pm \ImzSyst\syst \ .
\end{align*}

Since the SME fits consider only one specific frequency, \ie the sidereal
frequency, a wide range of frequencies is scanned by means of the periodogram
method. A periodogram gives the spectral power $P(\nu)$ of a frequency $\nu$ in
a signal sampled at discrete, not necessarily equidistant, times. In this
analysis, the Lomb-Scargle periodogram~\cite{Scargle:1982bw} is used, as in the
\babar measurement of \DamuBd~\cite{Aubert:2007bp}.

The periodogram is determined for the term in the decay rates proportional to
$e^{-\Gamma t}\Rez$. Since negative weights cannot be used in the periodogram,
the \B mass windows are narrowed to $5260 < m_{\jpsi\KS} < 5300\mevcc$ and $5350
< m_{\jpsi\Kp\Km} < 5390\mevcc$ compared to those used in the
fits~\cite{LHCb-PAPER-2015-004,LHCb-PAPER-2014-059}.  In total about 5200
frequencies are scanned in a wide range around the sidereal frequency, from 0.03
solar-day$^{-1}$ to 2.10 solar-day$^{-1}$. The number of frequencies oversamples
the number of independent frequencies by roughly a factor of two, thereby
avoiding any undersampling~\cite{Horne:1986bi}. As the data are unevenly
sampled, the false-alarm probability is determined from
simulation~\cite{Horne:1986bi}, where the time stamps are taken from data.

\begin{figure}
   \centering
   \includegraphics[width=\figWidth\textwidth]{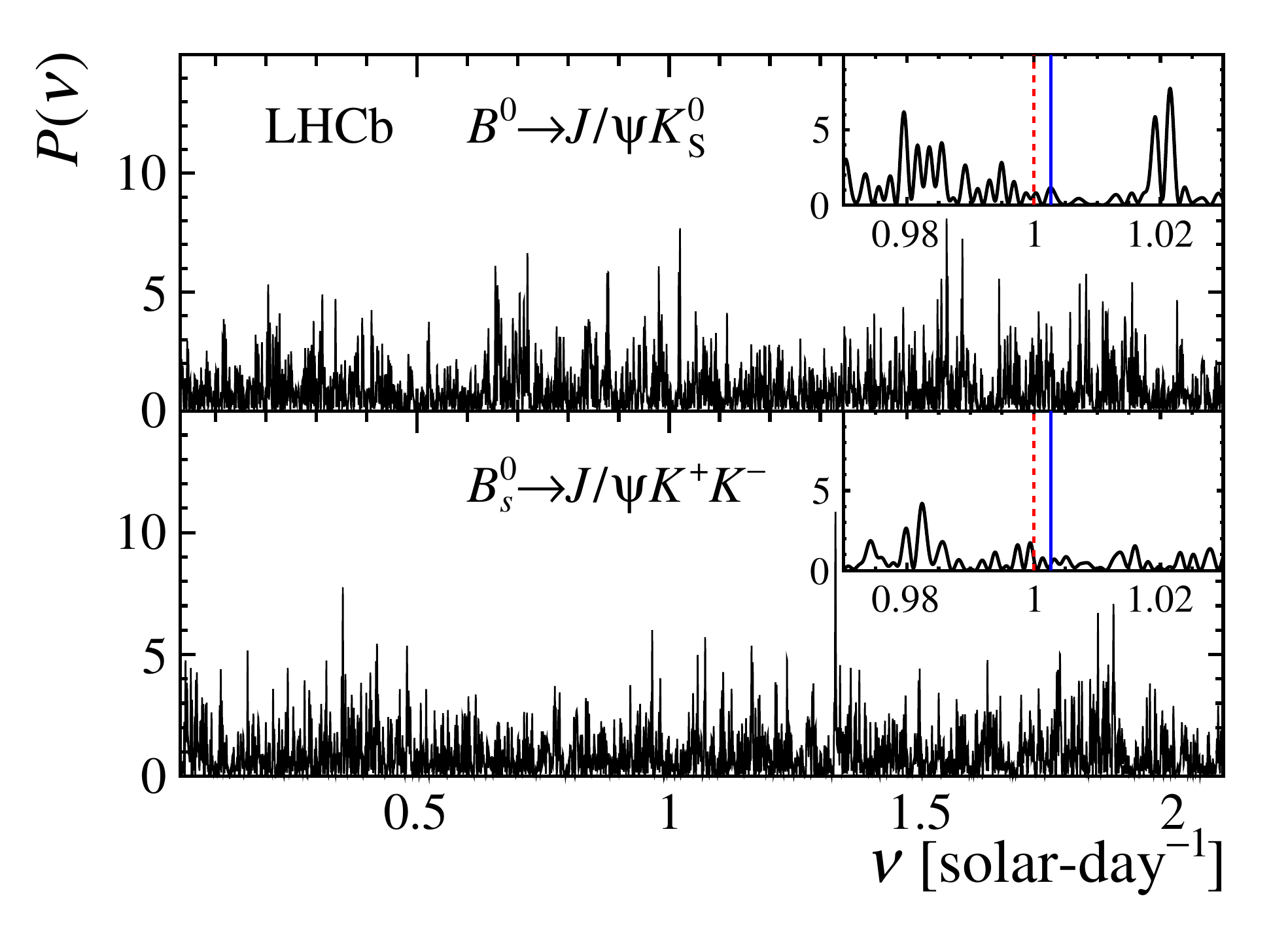}
   \caption{Periodograms for (top) the \Bd and (bottom) the \Bs sample. The
     insets show a zoom around the solar (red dashed line) and sidereal (blue
     solid line) frequencies, which have been made by highly oversampling the
     frequencies in this narrow range.}  \label{fig:periodogram}
\end{figure}

The two periodograms are shown in Fig.~\ref{fig:periodogram}. No significant
peaks are found. For the \Bd periodogram, the highest peak $P(\nu_{\rm max}) =
8.09$ is found at a frequency of 1.5507~solar-day$^{-1}$ and has a false-alarm
probability of 0.57.  There are 2707 (1559) sampled frequencies with a larger
spectral power than the peak at the sidereal (solar) frequency. For the \Bs
periodogram the highest peak $P(\nu_{\rm max}) = 10.85$ is found at a frequency
of 1.3301~solar-day$^{-1}$ and has a false-alarm probability of 0.06. There are
3386 (2356) frequencies with a larger spectral power than the sidereal (solar)
peak. The absence of any signal in the SME fits is confirmed by the absence of
significant peaks at the sidereal frequency.

The results presented here are consistent with \CPT symmetry and Lorentz
invariance. The measurement of \DamuBd is an improvement in precision of about
three orders of magnitude compared to the one from the \babar
collaboration~\cite{Aubert:2007bp} when the SM prediction $\DGd =
-0.0027\invps$~\cite{Lenz:2006hd} is used to scale their result. The measurement
of \DamuBs is an order of magnitude more precise than the one from the \dzero
collaboration~\cite{Abazov:2015ana} (note the different definition,
$\Dao\equiv\sqrt{\Dax^2+\Day^2}$, in Ref.~\cite{Abazov:2015ana}). The
measurement of \zBs is the first direct measurement of this quantity.

\section*{Acknowledgements}
\noindent We express our gratitude to our colleagues in the CERN
accelerator departments for the excellent performance of the LHC. We
thank the technical and administrative staff at the LHCb
institutes. We acknowledge support from CERN and from the national
agencies: CAPES, CNPq, FAPERJ and FINEP (Brazil); NSFC (China);
CNRS/IN2P3 (France); BMBF, DFG and MPG (Germany); INFN (Italy); 
FOM and NWO (The Netherlands); MNiSW and NCN (Poland); MEN/IFA (Romania); 
MinES and FANO (Russia); MinECo (Spain); SNSF and SER (Switzerland); 
NASU (Ukraine); STFC (United Kingdom); NSF (USA).
We acknowledge the computing resources that are provided by CERN, IN2P3 (France), KIT and DESY (Germany), INFN (Italy), SURF (The Netherlands), PIC (Spain), GridPP (United Kingdom), RRCKI and Yandex LLC (Russia), CSCS (Switzerland), IFIN-HH (Romania), CBPF (Brazil), PL-GRID (Poland) and OSC (USA). We are indebted to the communities behind the multiple open 
source software packages on which we depend.
Individual groups or members have received support from AvH Foundation (Germany),
EPLANET, Marie Sk\l{}odowska-Curie Actions and ERC (European Union), 
Conseil G\'{e}n\'{e}ral de Haute-Savoie, Labex ENIGMASS and OCEVU, 
R\'{e}gion Auvergne (France), RFBR and Yandex LLC (Russia), GVA, XuntaGal and GENCAT (Spain), Herchel Smith Fund, The Royal Society, Royal Commission for the Exhibition of 1851 and the Leverhulme Trust (United Kingdom).

\addcontentsline{toc}{section}{References}
\setboolean{inbibliography}{true}
\bibliographystyle{LHCb}
\bibliography{main,LHCb-PAPER,LHCb-DP,myrefs}

\newpage

\centerline{\large\bf LHCb collaboration}
\begin{flushleft}
\small
R.~Aaij$^{39}$, 
C.~Abell\'{a}n~Beteta$^{41}$, 
B.~Adeva$^{38}$, 
M.~Adinolfi$^{47}$, 
Z.~Ajaltouni$^{5}$, 
S.~Akar$^{6}$, 
J.~Albrecht$^{10}$, 
F.~Alessio$^{39}$, 
M.~Alexander$^{52}$, 
S.~Ali$^{42}$
% ...

{\footnotesize \it
% ...
$ ^{5}$Clermont Universit\'{e}, Universit\'{e} Blaise Pascal, CNRS/IN2P3, LPC, Clermont-Ferrand, France\\
$ ^{6}$CPPM, Aix-Marseille Universit\'{e}, CNRS/IN2P3, Marseille, France\\
% ...
$ ^{10}$Fakult\"{a}t Physik, Technische Universit\"{a}t Dortmund, Dortmund, Germany\\
% ...
$ ^{38}$Universidad de Santiago de Compostela, Santiago de Compostela, Spain\\
$ ^{39}$European Organization for Nuclear Research (CERN), Geneva, Switzerland\\
% ...
$ ^{41}$Physik-Institut, Universit\"{a}t Z\"{u}rich, Z\"{u}rich, Switzerland\\
$ ^{42}$Nikhef National Institute for Subatomic Physics, Amsterdam, The Netherlands\\
% ...
$ ^{47}$H.H. Wills Physics Laboratory, University of Bristol, Bristol, United Kingdom\\
% ...
$ ^{52}$School of Physics and Astronomy, University of Glasgow, Glasgow, United Kingdom\\
% ...
}
\end{flushleft}

\end{document}